\documentclass[aps,prl,reprint,groupedaddress]{revtex4-1}
\usepackage{graphicx}

\begin{document}


\title{Topological Solitons versus Non-Solitonic Phase Defects in Quasi-One-Dimensional Charge Density Wave}


\author{Tae-Hwan Kim}
\email[]{taehwan@postech.ac.kr}
\author{Han Woong Yeom}
\email[]{yeom@postech.ac.kr}
\affiliation{Center for Low Dimensional Electronic Symmetry and Department of Physics, Pohang University of Science and Technology, Pohang 790-784, Korea}

\date{\today}

\begin{abstract}
We investigated phase defects in a quasi-one-dimensional commensurate charge density wave (CDW) system, an In atomic wire array on Si(111), using low temperature scanning tunneling microscopy. The unique four-fold degeneracy of the CDW state leads to various phase defects, among which intrinsic solitons are clearly distinguished. The solitons exhibit a characteristic variation of the CDW amplitude with a coherence length of about 4~nm, as expected from the electronic structure, and a localized electronic state within the CDW gap. While most of the observed solitons are trapped by extrinsic defects, moving solitons are also identified and their novel interaction with extrinsic defects is disclosed. 
\end{abstract}


\maketitle


A topological soliton is a local solitary wave bridging two degenerate states, which has been an important conceptual tool in many branches of science~\cite{soliton2006}. For example, in electronic systems, solitons are responsible for the high electric conductivity in conjugated polymers such as polyacetylene~\cite{Su:1979fe,Su:1980pi}. A soliton exists as a nontrivial phase defect separating energetically degenerate one-dimensional (1D) charge density wave (CDW) states with an electronic state inside the CDW gap~\cite{Gorkov:1989cq,Brazovskii:2007uq,Brazovskii:2009qy}. 
While solitons in CDW systems show up their existence mainly through their consequences in transport and optical properties, it has been challenging to observe them directly in real space mainly because of their microscopic dimension and high mobility. This obstacle has prevented experimental investigations on various microscopic interactions of solitons predicted theoretically, such as those with other solitons and defects~\cite{Kivshar:1991il,Goodman:2004ee,Javidan:2006ye,Al-Alawi:2008gb,Hakimi:2009jl,Kartashov:2011zt}. 

The experimental difficulty can partly be overcome by the high spatial resolution of scanning tunneling microscopy (STM). Only very recently, a soliton was visualized by STM on an incommensurate 1D CDW system of NbSe$_3$~\cite{Brazovskii:2012kl}, which was immobilized by an unknown reason.  On the other hand, for a commensurate CDW system, a soliton has not been clearly identified partly because it can easily be confused with non-solitonic phase defects such as structural defects. Indeed, recent STM studies of metallic atomic wires on silicon surfaces in their CDW ground states revealed various phase defects~\cite{Morikawa:2004oz,Park:2004qa,Lee:2005rw,Snijders:2006dz,Zhang:2011hc}. However, the origin and solitonic characteristics of these phase defects have been debated and largely unclear~\cite{Yeom:2011dp,Zhang:2011tg}. 


In this Letter, we carefully reinvestigated with STM various local phase defects of In atomic wires on Si(111) in their quasi-1D CDW states~\cite{Morikawa:2004oz,Park:2004qa,Lee:2005rw,Zhang:2011hc}. We examine the solitonic characteristics of these defects such as their shapes (local variations of CDW envelopes), electronic states, and mobility. We can unambiguously distinguish topological solitons from various non-solitonic phase defects, solving the current debate on this system~\cite{Yeom:2011dp,Zhang:2011tg}. This opens up the possibility of investigating microscopic interactions of solitons in a commensurate CDW system and, in particular, the present work elucidates novel interactions of solitons with extrinsic phase defects.

The experiment was carried out with a ultrahigh-vacuum cryogenic STM (Unisoku, Japan). The Si(111)4$\times$1-In surface of In atomic wires was prepared by depositing 1~ML of In onto the clean Si(111)7$\times$7 surface at an elevated temperature~\cite{Yeom:1999fu}. Subsequently, the sample was cooled down to 78~K, well below the CDW transition temperature of $\sim125$~K, for STM measurements. All STM images presented here were taken in the constant-current mode with typically a tunneling current of 50~pA and a sample bias voltage of $-1$~V. 

\begin{figure}
\includegraphics[width=8cm]{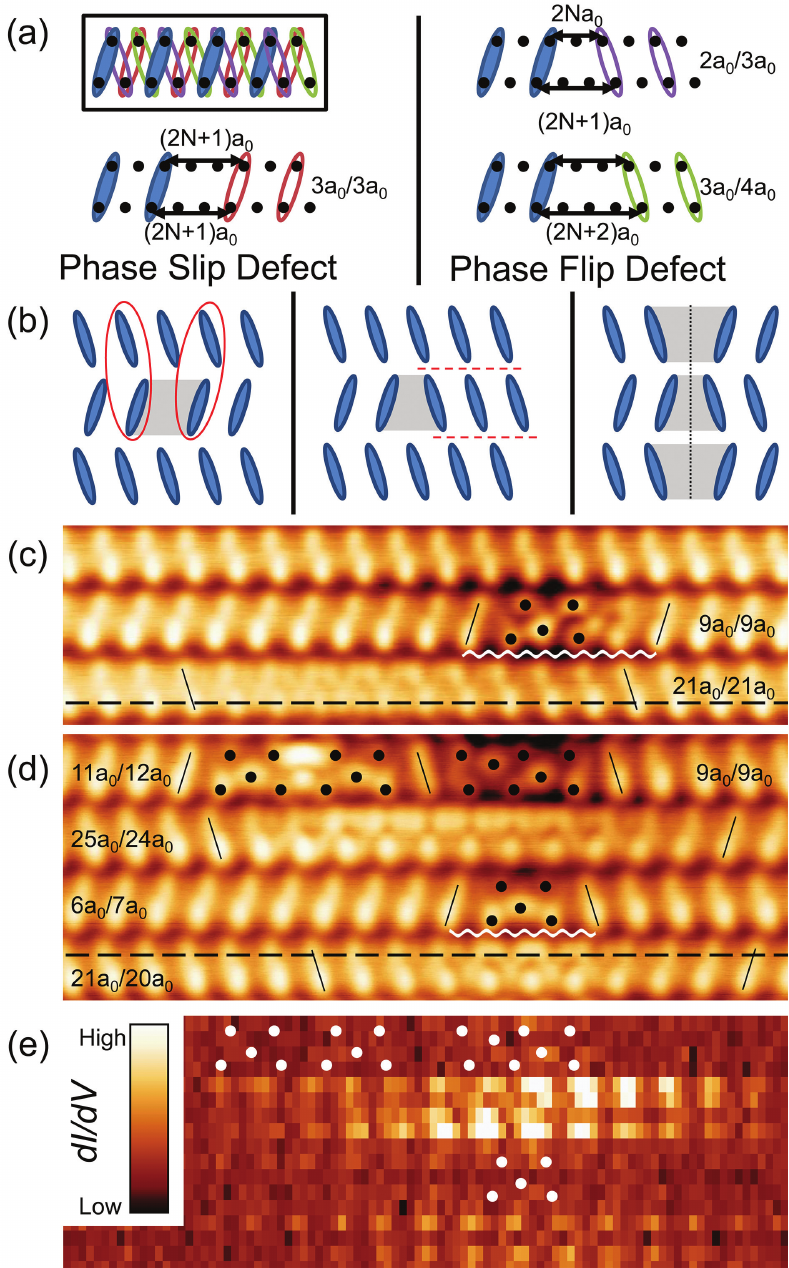}
\caption{(a) Schematic diagrams of three possible phase defects ($N=1$) on a 4$\times$2 CDW wire (solid ovals). Inset shows four degenerate CDW states in different colors for a single wire with two In atomic chains (black dots). The atomic structure is largely simplified~\cite{supplemental}. (b) Schematics of possible phase defects: a phase slip defect (PSD, left), a phase flip defect (PFD) with the broken 8$\times$2 order (middle), and a line boundary of PFDs (right). Two red ovals indicate two out of four possible 8$\times$2 configurations at the given CDW orientation. STM images showing (c) two PSDs and (d) PFDs and composite defects with (e) a corresponding differential conductance map taken at $+0.15$~V~\cite{supplemental}. Two nearest normal CDW maxima are used to determine lengths of defects as denoted by slashes. The distinct common local features of the defects are marked by dots [(c)--(e)].}
\end{figure}

An In wire of Si(111)4$\times$1-In has two In atomic chains and is separated by Si chains with periods of $4a_0$ and $1a_0$ perpendicular to and along the wire, respectively ($a_0$ is the Si lattice spacing, 0.384~nm)~\cite{Yeom:1999fu,supplemental}. Upon cooling, a commensurate 4$\times$2 CDW state develops through the periodicity-doubling distortion along a wire and shows a clear gap opening for a half-filled metallic band~\cite{Ahn:2004fk}. In addition, interwire coupling forces CDW wires to have opposite CDW orientations alternatively perpendicular to the wire, leading to 8$\times$2 ordering~\cite{supplemental}. Since there are four different ways to form a 4$\times$2 CDW state from two In atomic chains within a single wire [see the inset of Fig.~1(a)]~\cite{Hatta:2011ai}, we can expect up to three different kinds of phase defects (and their mirror symmetric versions) as described in Fig.~1(a). One is a phase \textit{slip} defect (PSD) between two-fold degenerate CDW states with the same CDW orientation, which are shifted by $\pi$ or half a CDW period ($1a_0$) along the wire [left panel in Fig.~1(b)]. On the other hand, two CDW states with opposite CDW orientations meet at a phase \textit{flip} defect (PFD). This defect is unconventional but comes from the unique double chain structure of the In wire; only one of two In atomic chains contains a $\pi$ phase shift. Unlike the PSD, the PFD lacks the two-fold degeneracy under the interwire coupling. Thus a PFD would induce two long domain walls  [dashed lines of middle panel in Fig.~1(b)] along the wire. Instead, the PFDs can exist near the edges of CDW wires such as step edges or cluster into a line boundary [right panel in Fig.~1(b)] of two CDW domains across the wires while various isolated PSDs are observed~\cite{supplemental}.

Figures~1(c) and 1(d) show STM images of various phase defects, which are static enough to be imaged by STM. Two PSDs in Fig.~1(c) show contrasting length scales. A short PSD has a distinct local structure as indicated by dots [Fig.~1(c)], which abruptly induces the CDW phase shift. In contrast, a long PSD in the neighboring wire gradually changes the CDW phase over a relatively long distance of $\sim20a_0$ without such a distinct local structure. The latter length is consistent for most of the long PSDs observed. On the other hand, as mentioned above, the PFDs tend to form a line boundary perpendicular to the wires [Fig.~1(d)]~\cite{supplemental}. The short PFD has a similar local structure [dots in Fig.~1(d)] to a short PSD. A long PFD changes its phase gradually like a long PSD but for only one In chain in the wire. The short PFDs as well as the short PSDs seem to act as building blocks to form various composite phase defects as shown in the top wire of Fig.~1(d). 

\begin{figure}
\includegraphics[width=8cm]{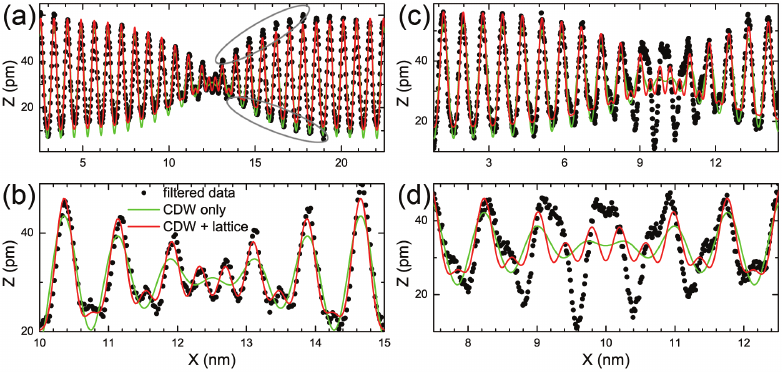}
\caption{STM line profiles (dots) of (a) a long PSD and (c) a long PFD obtained along the dashed lines in Figs.~1(c) and 1(d) together with their best fits by the soliton shape with (red) and without (green) considering the lattice modulation. The line profiles are high-pass filtered to get rid of slowly varying backgrounds. With the lattice modulation, the soliton shape is modified as $Z^*_S(x) = Z_S(x) +A_L \cos[2\pi (x-x_0)/a_0]$, where $A_L$ is the height modulation of underlying atoms. The inclusion of the lattice modulation results in a much better fit with almost the same coherence length. $\xi_{\rm (a)} = 3.75 \pm 0.14$ ($3.75 \pm 0.17$)~nm and $\xi_{\rm (c)} =3.50 \pm 0.23$ ($3.40\pm0.26$)~nm with (without) the lattice modulation. (b) and (d) Enlargements of (a) and (b) around the soliton center.}
\end{figure}

The long PSDs are similar to what was assigned as a soliton by Morikawa \textit{et al.}~\cite{Morikawa:2004oz} and the short PFDs were argued as solitons by Zhang \textit{et al}.\ mainly due to the phase shift involved~\cite{Zhang:2011hc,Yeom:2011dp,Zhang:2011tg}. Note, however, that various extrinsic (structural, chemical, and charge) defects can in principle cause CDW phase shifts, which should be distinguished from a intrinsic soliton as the lowest-energy excitation from the CDW ground state~\cite{Brazovskii:2012kl}. In terms of the shape, such a soliton excitation would have a well defined shape $Z_S(x) = Z_0 +A_C \tanh [(x-x_0)/\xi] \sin[\pi (x-x_0)/a_0]$, where $A_C$ is the CDW amplitude and $\xi$ is the microscopic coherence length~\cite{Gorkov:1989cq,Brazovskii:2012kl}. The CDW amplitude varies smoothly within the coherence length, half the apparent size of a soliton, which is determined by $\xi = \hbar v_F /\Delta$, where $v_F$ is the Fermi velocity and $\Delta$ is half the CDW gap~\cite{Su:1980pi,Nakahara:1990rq}. For the present case, the band dispersions~\cite{Ahn:2004fk} and the gap size~\cite{supplemental} yield  $\xi = 3.4$~nm $\sim 9a_0$, which is similar to that of the soliton in polyacetylene, $\sim7a$~\cite{Su:1979fe}. It is thus apparent that the short PSDs and short PFDs are not solitons judging from their shapes and lengths ($<2\xi$$\sim$$18a_0$). 
They are most likely structural defects or adsorbate-induced ones. Structural defects with a $\pi$ phase shift in the dimerized chain structure were observed in various systems including the clean Si(001)2$\times$1~\cite{Pennec:2006pt} and Si(111)5$\times$2-Au surfaces~\cite{Kang:2008fv}.

However, the long phase defects are clearly different. We fitted a line profile obtained from a long PSD [Fig.~1(c)] with the above soliton shape. As shown in Fig.~2(a), the profile is excellently fitted; the gradual decay of the CDW amplitude and the $\pi$ phase shift around the center are clearly visualized. The coherence length fitted is $\xi = 3.75 \pm 0.17$~nm [Fig.~2(a)] in good agreement with that estimated above. This form is very much consistent with the trapped soliton observed in NbSe$_3$~\cite{Brazovskii:2012kl}. We thus interpret that a long PSD is an intrinsic soliton immobilized by some trapping potential. The trapping potential is obviously provided by a neighboring short PSD [indicated by the wavy line in Fig.~1(c)] since most of long PSDs are observed as paired with neighboring defects. This trapping potential affects the soliton shape to deviate marginally from its theoretical form as indicated by the ovals in Fig.~2(a). The deformation is more significant on the In atomic chain closer to the neighboring defect~\cite{supplemental}. 

A profile of a long PFD can also be consistently fitted with the same function with a similar coherence length [Fig.~2(c)]~\cite{supplemental}. It suggests that the long PFDs are also similar solitons. However, the deviation near a neighboring defect, which is located near the center, is apparently larger in this case [Fig.~2(d)].  This may be explained by a CDW perturbation due possibly to the Friedel oscillations induced by a neighboring defect [the wavy line in Fig.~1(d)] as also observed in NbSe$_3$~\cite{Brazovskii:2012kl}. Including the local electronic response to the neighboring defect, we could reproduce precisely the shape of a PFD with the soliton profile~\cite{supplemental}.


Since a soliton would have an electronic state within the CDW gap~\cite{Gorkov:1989cq,soliton2006,Brazovskii:2007uq,Brazovskii:2009qy}, we checked this localized electronic state by scanning tunneling spectroscopy for short and long PFDs~\cite{supplemental}. From the differential tunneling conductance map obtained just below the upper CDW gap edge [Fig.~1(e)], we clearly found that the long PFD shows a localized state ($\sim+0.15$~eV) within the band gap with a similar length to the coherence length in contrast to the short PFDs and their composite defects~\cite{supplemental}. The deviation of this soliton state from the Fermi energy may be ascribed to the structural or electronic perturbation on the long PFD by the neighboring defect or a strong interaction with the underlying lattice in a commensurate CDW system, where the CDW phase is completely locked to the underlying lattice. On the other hand, a similar spectroscopy measurement on a long PSD was hardly possible since it can rather easily  be detrapped  during a long measurement time for spectroscopy. 


\begin{figure}
\includegraphics[width=8cm]{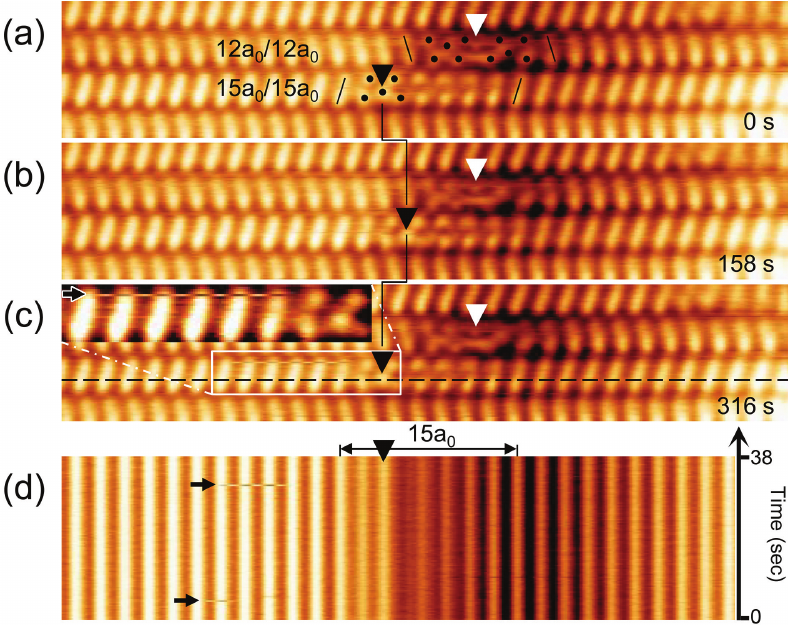}
\caption{(a)--(c) A series of STM images taken consecutively with their start times indicated. Two defects indicated by white and black triangles are a reference and a hopping PSD, respectively. Sometimes, two abrupt CDW phase shifts ($0 \rightarrow \pi \rightarrow 2\pi$) occur in a short time span as marked by a black arrow in the inset of (c). (d) Stacked STM line profiles (the newest line profile at top most) taken along a dashed line in (c). Each black arrow indicates the two successive CDW phase shifts.}
\end{figure}


The trapping or detrapping assumes another important characteristic of a soliton to be addressed, the mobility. Therefore, the assignment of a trapped soliton would not be complete without observing freely-moving solitons. In order to find any moving solitons, we scanned the same restricted area repeatedly. Figures~3(a)--(c) are a series of sequential STM images showing the same CDW wires around two composite defects with (a $15a_0$-PSD) and without ($12a_0$) a phase shift. One can observe that the $15a_0$-PSD hops by $2a_0$ to the right [Fig.~3(b)] and then hops back to the original position in a time scale of a few minutes [Fig.~3(c)]. Most of the short defects showed a similar random hopping motion by a multiple of a CDW period ($2a_0$) at a time. This motion is quite similar to that previously reported by Zhang and others~\cite{Zhang:2011hc}. However, such a hopping motion is natural for trivial defects like the non-solitonic short PSDs. 

In sharp contrast to the low frequency hopping, we could observe a much faster (beyond our temporal resolution) shift of a CDW phase as marked by the black arrow in the inset of Fig.~3(c). A $\pi$ phase shift occurs abruptly but the phase returns back in a short time of $\sim50$~msec. This abrupt phase shift is more clearly visualized by continuously scanning the same line along one CDW wire. Figure~3(d) shows an example of such a line scan series, where two events of abrupt phase shifts are observed (indicated by black arrows). Each event consists of two successive $\pi$ phase shifts ($0 \rightarrow \pi \rightarrow 2\pi$). These fast phase shifts may result from a moving soliton that moves too fast to be captured with STM. Then, the above STM image would indicate two solitons passing sequentially the scanning tip. Rather frequently, these abrupt CDW phase shifts occur together with the hopping of the PSD as shown in Fig.~4(a). For a more detailed analysis of these complex motions, we extract individual line profiles from Fig.~4(a), especially, at close range of the hopping of the PSD. We found that the PSD (black triangles) hopped by $1a_0$ right after an abrupt $\pi$ phase shift (black arrows) and two successive phase shifts induced the PSD to hop by $2a_0$ in total between the scan (0) and (4) [Fig.~4(b)]. The same successive but reverse hoppings occurred between the scan (7) and (11) with two abrupt phase shifts. 

\begin{figure}
\includegraphics[width=8cm]{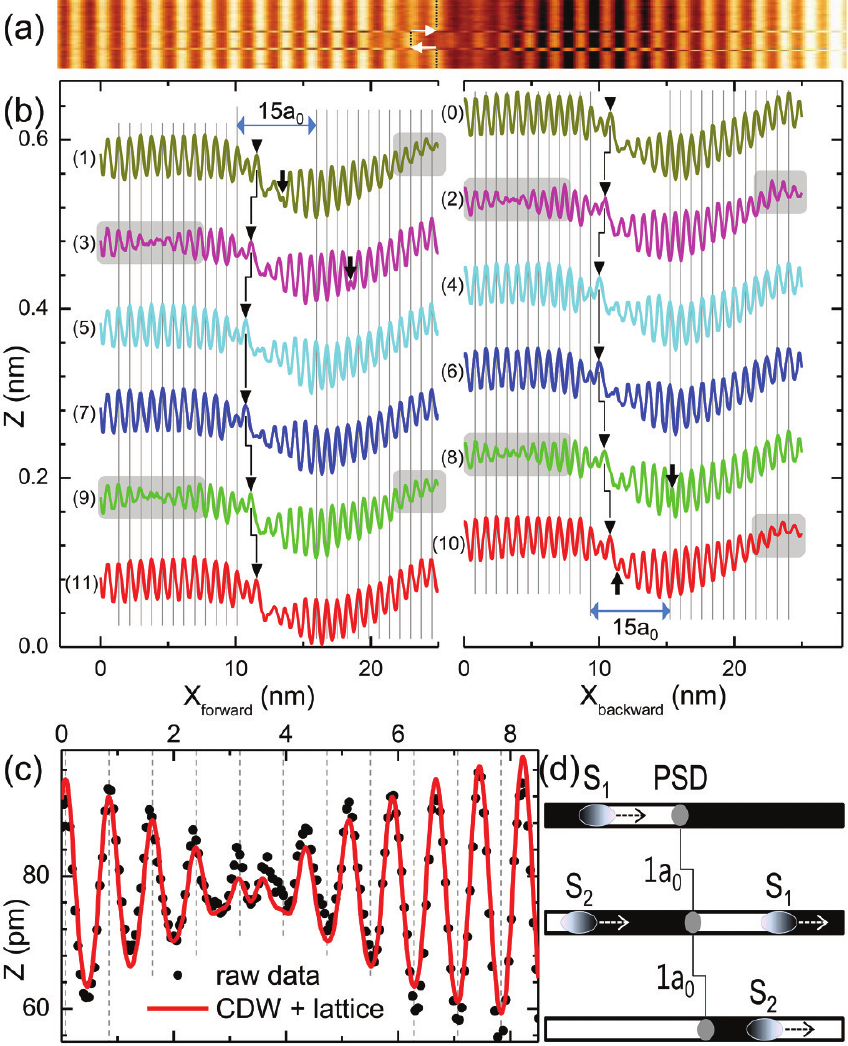}
\caption{(a) Stacked STM line profiles showing the hopping of the PSD. Dotted lines and white arrows indicate the positions of the PSD and the hopping directions, respectively. (b) The STM profiles extracted from (a) are displayed in a time sequence, with the earliest on the top (0) and the latest on the bottom (11). Left (right) curves show the line profiles obtained by scanning from left (right) to right (left). The scan speed was set to 150~msec per line. Black triangles and vertical arrows indicate the positions of the PSD and abrupt CDW phase shifts, respectively. Gray boxes highlight emerging long PSDs or solitons. (c) Fitting of a soliton's profile (dots) with a soliton solution (solid line, $\xi = 3.56 \pm 0.48$~nm). Note that raw data was used for fitting in contrast to the trapped solitons. (d) Schematics of the interaction between solitons ($S_{1,2}$) and a PSD. Black and white areas indicate different CDW states.
}
\end{figure}



More interestingly, we observed a few long PSDs transiently [indicated by gray boxes in Fig.~4(b)]. The detailed line profile [Fig.~4(c)] clearly indicates that they are the same as the static solitons discussed above. As a rule, they appeared after an abrupt $\pi$ phase shift [$0 \rightarrow \pi$, see the scan (1) and (2)] and disappeared after another $\pi$ phase shift [$\pi \rightarrow 2\pi$, (3) and (4)]. This indicates that these PSDs are moving solitons trapped transiently and the abrupt CDW phase shifts are created when these solitons are trapped or detrapped. 

The transient trapping of a moving soliton can be explained by the soliton-barrier model. Theoretical simulations show that the scattering of a soliton with a potential barrier is nearly elastic~\cite{Javidan:2006ye,Al-Alawi:2008gb,Hakimi:2009jl}. For an initial velocity $v_i $ smaller (larger) than the critical velocity $v_c$, a soliton reflects back (transmits over the barrier) [Fig.~3(d)]. Interestingly, at $v_i \approx v_c$, a soliton interacts with the barrier slowly and stays for a while near the barrier until it finally escapes. If the present soliton at a velocity close to $v_c$ interacts with the potential barrier of a short PSD, the observed motion can be explained qualitatively well. One thing not captured in this theory is that in our case the potential barrier itself moves when a soliton transmits through it [Fig.~4(d)]. It is natural that a soliton with a $\pi$ phase shift in itself translates a static phase defect by $\pi$ or $1a_0$. Since the soliton motion and the hopping of extrinsic defects are coupled, it might require an extra energy cost for a soliton transmission. 

It is noteworthy that most of the hopping events observed are 2a$_0$-hoppings consisting of two $1a_0$-hoppings in a very short time span [Fig.~4(d)]. This suggests that two solitons, that is, a soliton and an antisoliton, are combined to move as predicted in theory~\cite{soliton2006}. This needs further investigation along with the issue of separating pure thermal hoppings of the defects from the soliton-induced ones.

This work was supported by National Research Foundation of Korea (NRF) through Center for Low Dimensional Electronic Symmetry (Grant No. 2012R1A3A2026380) and SRC Center for Topological Matter (Grant No. 2011-0030789).

\bibliography{soliton}

\end{document}